# Molecular Electronics:
# electron, spin and thermal transport through molecules.

Dominique Vuillaume

*Institute of Electronic Microelectronics and Nanotechnology (IEMN), CNRS*
*Avenue Poincaré, Villeneuve d'Ascq (France).*

## 1. Introduction

It is widely accepted that molecular electronics (ME) started in 1971 when Mann and Kuhn were the first to measure electron transport and demonstrate tunneling effect through a monolayer of organic molecules (alkyl chains) connected between two electrodes.[1] In 1974, Aviram and Ratner went a step further when they theoretically proposed the concept of a molecular rectifier (molecular diode).[2] They introduced the idea that a donor-acceptor molecule can rectify the current as in the p-n semiconductor junctions. We nevertheless note that the foundations of these ideas are the former works by R.S. Mulliken who developed the concept of molecular orbitals (in the 1940s) and later made seminal works on charge transfers in donor-acceptor complexes.[3] Another seminal contribution, namely that protein molecules can conduct electricity, was proposed by A. Szent-Györgyi in the same period.[4] The reader interested in an historical review on the foundations of ME is referred to [5].

The Aviram-Ratner molecular rectifier was experimentally demonstrated in 1990 by Ashwell *et al.*[6,7] and then extensively studied by many groups, especially by Metzger *et al.*[8-10], the reader is referred to an extended review published recently.[11] Nowadays, due to molecular engineering and optimization of the molecule/electrode interactions, the performances of the molecular rectifiers, i.e., the rectification ratios (ratio of the current at two opposite voltage polarities) are in excess of $10^5$ on a par with their semiconductor counterparts.[12]

ME aims at "information processing using photo-, electro-, iono-, magneto-, thermo-, mechanico or chemio-active effects at the scale of structurally and functionally organized molecular architectures" (adapted from Ref. 13). ME is of great interest for basic science at the nanoscale because molecules are quantum objects by nature, chemists can tailor their properties at the synthesis level opening many perspectives for new experiments.

This review presents recent results on the physics of electron transport in molecular devices. The review is organized as follows. A brief description of molecular junction (MJ) technology is first given followed by an introduction to the basic physics of electron transport through MJs from DC to ~20 GHz. Then, several sections review selected results on spin-dependent transport, plasmonics, quantum interferences, thermal transport and electronic noise in ME devices.

## 2. How to make a molecular junction

The term molecular junction (MJ) refers to a simple device where molecules (from a monolayer to a single molecule) are connected between two electrodes (metal or semiconductor), Fig. 1. The number of molecules in the MJ depends on the size of these electrodes and the MJs are usually classified as "large area MJ" and "single (or a few) MJ". Albeit the Langmuir-Blodgett method was used in the early times of ME to deposit molecules on the electrode surfaces (only large area MJs), chemisorption is nowadays widely used because it is also suitable at the nanoscale to attach few molecules between electrodes of a nanometer dimension. In that case, the molecules are equipped with anchor groups at both (or only one) ends. The chemical nature of the anchor group is chosen depending on the nature of the electrodes to permit a chemical reaction with the electrode by forming a chemical bond between the molecule and the electrode. Archetypes of anchor groups are thiols (-SH) on metals (Au, Ag,…), alkenes (-C=C-) on hydrogenated-silicon surfaces, a detailed review is given in Refs [14-16]. This approach is widely versatile to adapt the molecules on the electrodes of interest. However, this chemical link also has a pronounced effect on the global electronic properties of the MJs, mainly governing the electronic coupling between molecules and electrodes (see below in this chapter) and this point has been extensively studied (see a review [17]). For large are MJs (left part of Fig. 1), in a vertical configuration, the top electrodes (typically with a lateral size > few μm, roughly speaking ≳ $10^8$ molecules in the MJs) can be fabricated by metal evaporation. However, due to the extreme thickness of the molecular layer (monolayer), the metal atoms can easily diffuse into the monolayers resulting in nanoscale metallic shorts and a low yield of fabrication. This drawback can be avoided by intercalating a thin layer of a highly conductive polymer ("organic metal") acting as a protective buffer between the monolayer and the metal electrode [18] and soft metal transfer techniques have been developed. [19-21] The other solution, most widely used for laboratory experiments, is the use of soft metal contacts such as mercury drops [1,22] or eutectic GaIn drop contacts [23], the latter being combined with microfluidic systems for a better control and stability.[24]

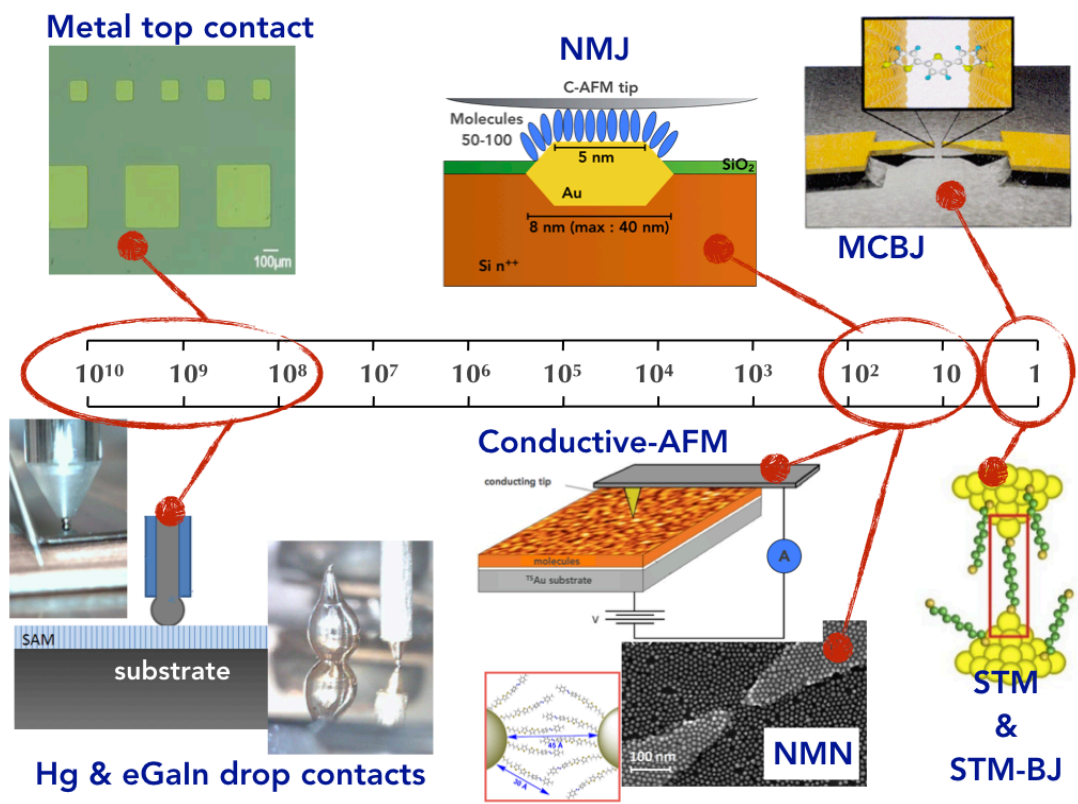

*Figure 1.* Overview of several types of molecular junctions (MJs) along a scale of the approximate number of molecules in the MJ: from "large area MJ" on the left to few molecules and single molecule junction on the right.

At the nanoscale, scanning tunneling microscope (STM) and conductive probe atomic force microscope (C-AFM), are the tools of choice to study a single or a few (typically ≲ 100) molecules (right part of Fig. 1). For single molecule experiments, STM break-junctions (STM-BJ) [25] and mechanically controlled break junctions (MCBJ) [26] have been developed (see a review [27,28]). In these BJ approaches, the two electrodes are repeatedly (up to thousands or more) moved close and apart and a molecule bridges the electrode gap when its size matches the electrode gap. The signature of the molecule conductance appears as plateaus in the current vs. gap electrode distance traces. C-AFM is used to gently contact (weak loading force) the monolayer [29-31] on large surfaces and to measure the current voltage at few hundreds places on the monolayer. Another approach is the use of tiny nanodot electrodes (nanodot-molecule junction, NMJ).[32,33] In this latter case, hundreds to thousands of nanodots (typically 5-40 nm in diameter) are fabricated by e-beam lithography and can be measured in a "one shot" single C-AFM image. Note that all these techniques require a large number of measurements and a solid statistic analysis to get reliable conductance values of the MJs. Finally, another approach is to use a 2D network of metal nanoparticles (NP)-molecule network (NMN).[34-37] The basic element is a NP-molecules-NP junction with few molecules bridging the gap between two adjacent NPs (typically around 10 nm in diameter), its conductance being extracted from the global conductance of the NMN, if the NP-molecule-NP elements are reasonably ordered and the network structure known (e.g., via scanning electron microscopy).

More details on these technological issues can be found in [38-40].

### 3. Electron transport in molecular devices: back to basics

We briefly expose how electrical charges (electrons and holes) are transported through a molecule (or an ensemble of molecules, *e.g.*, a molecular monolayer) connected between two electrodes. For more details, the reader is referred to text books [41, 42].

Figure 2 shows the energy diagram of a molecular junction. In this simplified scheme, the electrodes are modeled by their Fermi energy, $\varepsilon_F$, and the molecule by their quantized energy levels (molecular orbital, MO) of which the highest occupied molecular orbital (HOMO, $\varepsilon_H$) and the lowest unoccupied molecular orbital (LUMO, $\varepsilon_L$) are playing the most role in the charge transport. All energies are referred to the vacuum energy level. Two other parameters are required, $\Gamma_L$ and $\Gamma_R$, to model the interactions of the molecules with the left and right electrodes (Fig. 2a). These energies correspond to the interaction between the electrons in the molecules and the electron cloud in the electrodes, *i.e.*, the coupling between the molecular orbitals and the density of states in the electrodes. As a result of this hybridization of the MOs with the delocalized states of electrons in the electrodes, the MOs are broadened by a quantity $\Gamma = \Gamma_L + \Gamma_R$. This coupling energy, typically ~ 0.1 to 10 meV, mainly dictates the physical nature of the electron transport. In the weak coupling regime (~ 0.1 meV), the molecule can be considered as a quantum dot with an ultra-thin tunnel barrier at the molecule/electrode interface. More precisely, this regime corresponds to $\Gamma << U$, with U the additional energy, i.e., the energy required to add an electron to the LUMO ($U_{N+1}-U_N$) or to remove one electron from the HOMO ($U_{N-1}-U_N$), $U_N$ being the total energy of the system with N electrons. In this regime, Coulomb blockade (at low enough temperature), incoherent transport (the electron wave function can be perturbed during the transport) are usually observed. This regime (Coulomb blockade) is not discussed in this chapter, for more details see [41, 43]. In the medium-strong coupling regime (considered here), the transport is usually coherent and can be classified as off-resonant transport (Fig. 2b) or resonant transport, Fig. 2c (see next section). The coupling energy is determined by the physical or chemical nature of the molecule/electrode interface. The weak coupling case generally corresponds to physisorbed molecules or molecules chemisorbed on electrodes covered by a very-thin insulating layer (e.g., native oxide on silicon or on Al, metal covered by few atomic layers of insulating materials like NaCl or KBr on Ag for instance). For anchoring molecules on metal (e.g., Au), molecules equipped with thiol anchoring groups at their ends are usually used giving a larger coupling energy, but other anchoring groups have also been studied (e.g., amine, pyridine, cyano, carboxyl), in each case, the coupling being strongly dependent on the atomic detail configuration between the anchoring group and the electrode surface (see reviews in [17, 44-46]). In both coupling regimes, and whatever the transport is coherent or incoherent, we also distinguish elastic (at constant energy) and inelastic electron transport (Fig. 2d), where the energy of the electron is modified by various interactions (e.g., with the vibration modes of the chemical bonds in the molecule).

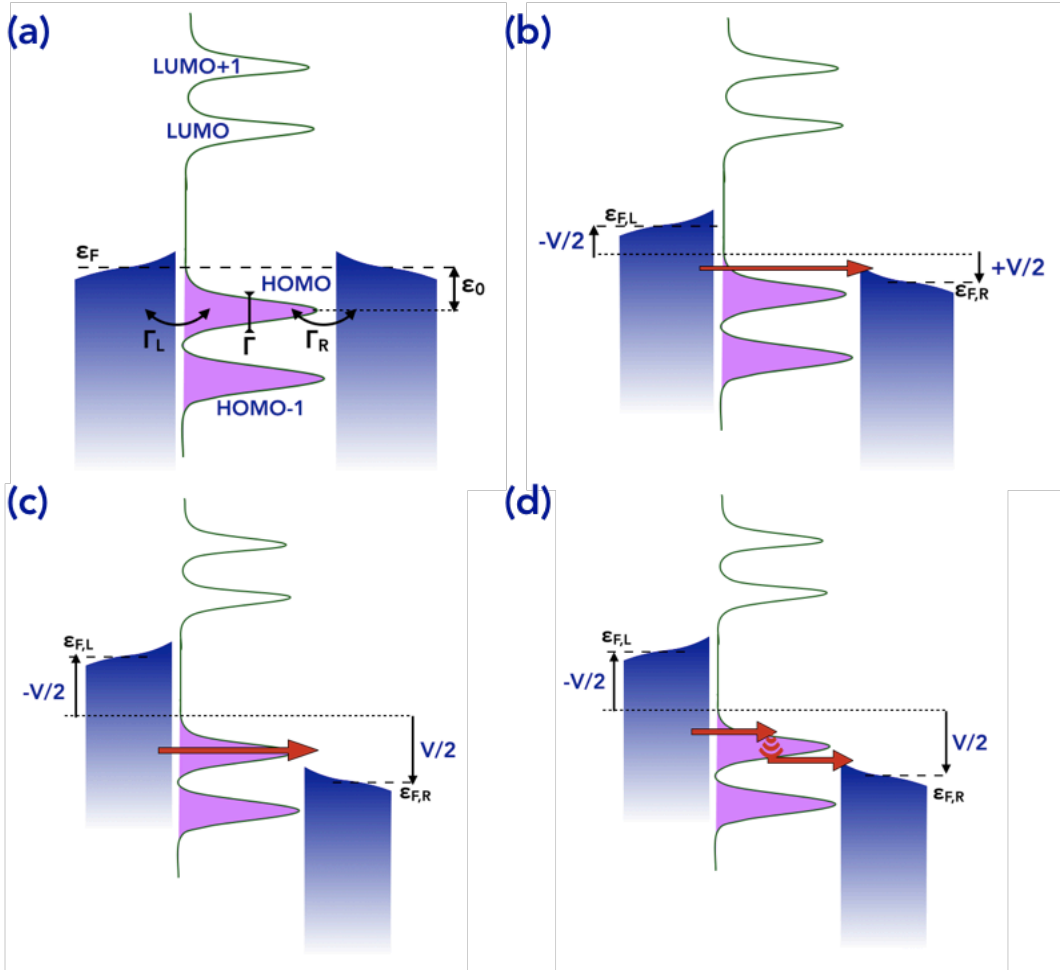

*Figure 2.* (a) Simplified energy diagram of a molecular junction. (b) Coherent, off-resonant, electron transport, eV < 2ε₀. (c) Coherent, resonant, electron transport, eV ≥ 2ε₀. (d) Inelastic electron transport.

## 4. Electron transport: DC and low frequency

At low voltages (Fig. 2b) the quantum tunneling effect is the dominant electron transport mechanism. At higher voltages (Fig. 2c), resonant transport through one of the molecular orbitals (here HOMO for illustration) is allowed as soon as eV=±2$\varepsilon_0$, with $\varepsilon_0$ the energy position of the involved MO with respect to the electrode Fermi energy. The key characteristic of such a molecular junction is its current versus voltage, I(V), curve. The I(V) characteristic is described by the Büttiker-Imry-Landauer formalism:[47]

$$I(V) = \frac{2e}{h}\int T(E)\left[f(E,\varepsilon_{F,L}) - f(E,\varepsilon_{F,R})\right]dE \qquad [1]$$

where T(E) is the transmission coefficient (electron transmission probability) through the molecule, f the Fermi-Dirac statistics $f(E,\varepsilon) = \left[1 + \exp\frac{E-\varepsilon}{k_B T}\right]^{-1}$, e the electron charge, $k_B$ the Boltzmann constant, T the temperature, h the Planck constant and $|\varepsilon_{F,L} - \varepsilon_{F,R}| = eV$, V the applied voltage. T(E) is determined using first principal calculations, such as Non-Equilibrium Green Functions (NEGF) combined with Density Functional Theory (DFT), see [41]. Figure 3 shows such a calculated T(E) and the corresponding I(V) curves for an azobenzene derivative molecule contacted between Co and Au electrodes and for several conformation of the molecule in the molecular junction.[48] If we consider that that there is no peak of T(E) near $\varepsilon_F$ (i.e., off-resonance, $\varepsilon_F$ far enough from the LUMO and HOMO levels) and low temperature the conductance of the molecular junction can be simplified (from Eq. [1]) to:

$$G = \frac{2e^2}{h} T(\varepsilon_F) = G_0 T(\varepsilon_F) \qquad [2]$$

with G0 the quantum of conductance (77.5 μS). A simple analytical model can be derived if we consider that i) a single MO (either LUMO or HOMO) dominates the charge transport, ii) that the voltage mainly drops at the molecule/electrode interface (i.e., no shift of the molecular orbital with the applied voltage) and iii) that the MO broadening is described by a Lorentzian or Breit-Wigner distribution.[41, 49] Then, the transmission coefficient can be simplified and writes:

$$T(E) = \frac{4\Gamma_L \Gamma_R}{(E-\varepsilon_0)^2 + (\Gamma_L + \Gamma_R)^2} \qquad [3]$$

In this single energy-level model, at low temperature, we get the following analytical expression[49]

$$I(V) = \frac{8e}{h} \frac{\Gamma_L \Gamma_R}{\Gamma_L + \Gamma_R} \left( \arctan\left( \frac{\varepsilon_0 + \frac{\Gamma_L}{\Gamma_L + \Gamma_R} eV}{\Gamma_L + \Gamma_R} \right) - \arctan\left( \frac{\varepsilon_0 - \frac{\Gamma_R}{\Gamma_L + \Gamma_R} eV}{\Gamma_L + \Gamma_R} \right) \right) \qquad [4]$$

with $\varepsilon_0 = \varepsilon_{L/H} - \varepsilon_F$ depending on the MO involved in the transport. This equation is used to fit the experimental data and to extract the parameters of the model $\varepsilon_0$, $\Gamma_L$ and $\Gamma_R$.

A low bias (off-resonant transport), this equation gives an S-like shape I(V) curve with a step-like increase of the current (in absolute value) when approaching the resonant transport condition and a plateau for voltages larger than V=2$\varepsilon_0$/e (in absolute value). Figure 4 shows simulated I(V) curves with various values of the parameters $\varepsilon_0$, $\Gamma_L$ and $\Gamma_R$. The parameter $\varepsilon_0$ mainly control the position of the resonant transport step and the curvature of the S-like shape at low bias (V<2$\varepsilon_0$/e) – Fig. 4a, while the coupling energy $\Gamma_L$ and $\Gamma_R$ dictate the current amplitude (Fig. 4b). Asymmetric I(V) curves (rectification diode behavior) are obtained with $\Gamma_L \neq \Gamma_R$, a situation experimentally encountered when the MOs involved in the electron transport are geometrically located in an asymmetric position in the junction, e.g., when the molecule is closer to an electrode than the other,[12, 50, 51] or when the molecule is coupled to the electrodes via two chemically different anchoring group (Fig. 5).[52] Note that this single energy-level model has to be used with caution. Eq. 4 is a low temperature approximation and it can be used at room temperature for voltages below the resonant transport conditions [52, 53] since the temperature broadening of the Fermi function is not taken into account.

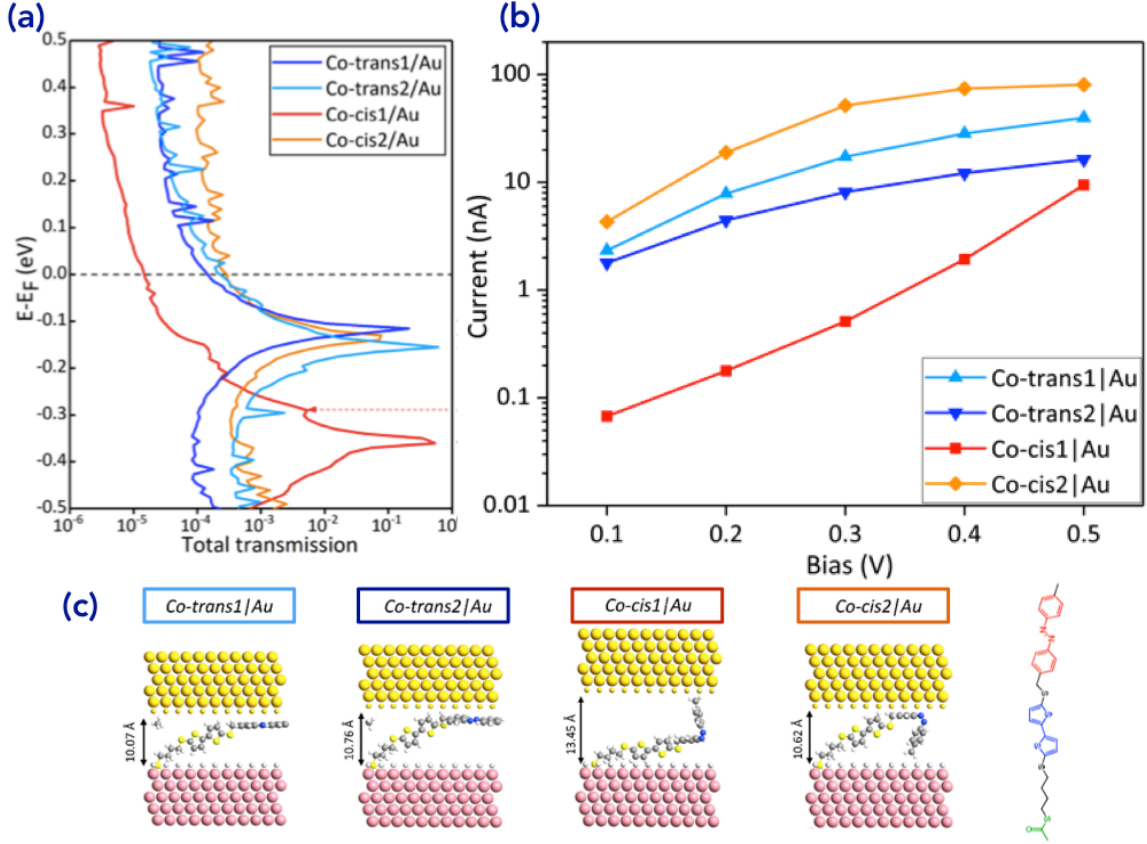

*Figure 3.* (a) Calculated (DFT and NEGF) electron transmission coefficient T(E) and (b) corresponding I(V) curves using Eq. 1 for a Co/molecule/Au junction. (c) Calculated optimized geometry for 4 conformations of the azobenzene-bithiophene-alkylthiol molecule (chemical structure shown at the right side).[48]

Similarly, due to its simplicity the same I(V) curve can account for two different situations: an electron transport through the HOMO strongly coupled to the right electrode (red curve in Fig. 4c) and an electron transport though the LUMO strongly coupled to the left electrode (green squares in Fig. 4c). Finally, this model does not capture some important effects like the fact that the MO energy level in the junctions and the coupling parameters are dependent on the applied voltage. Recently, it has been demonstrated that fixing the low-bias conductance value and the Seebeck coefficient in the fit protocol improve the determination of the MO energy level.[54] Another approach to measure $\varepsilon_0$ is to determine the voltage at which it is aligned with the Fermi energy of one of the electrodes. In this technique, known as transition voltage spectroscopy (TVS), plotting abs(V²/I) versus V gives two peaks (positive and negative threshold voltage : $V_{T+}$ and $V_{T-}$) when the levels are aligned (Fig. 4d).[55-59] The MO energy level is given by:[57]

$$|\varepsilon_0| = 2 \frac{e|V_{T+}V_{T-}|}{\sqrt{V_{T+}^2 + 10|V_{T+}V_{T-}|/3 + V_{T-}^2}} \quad [5]$$

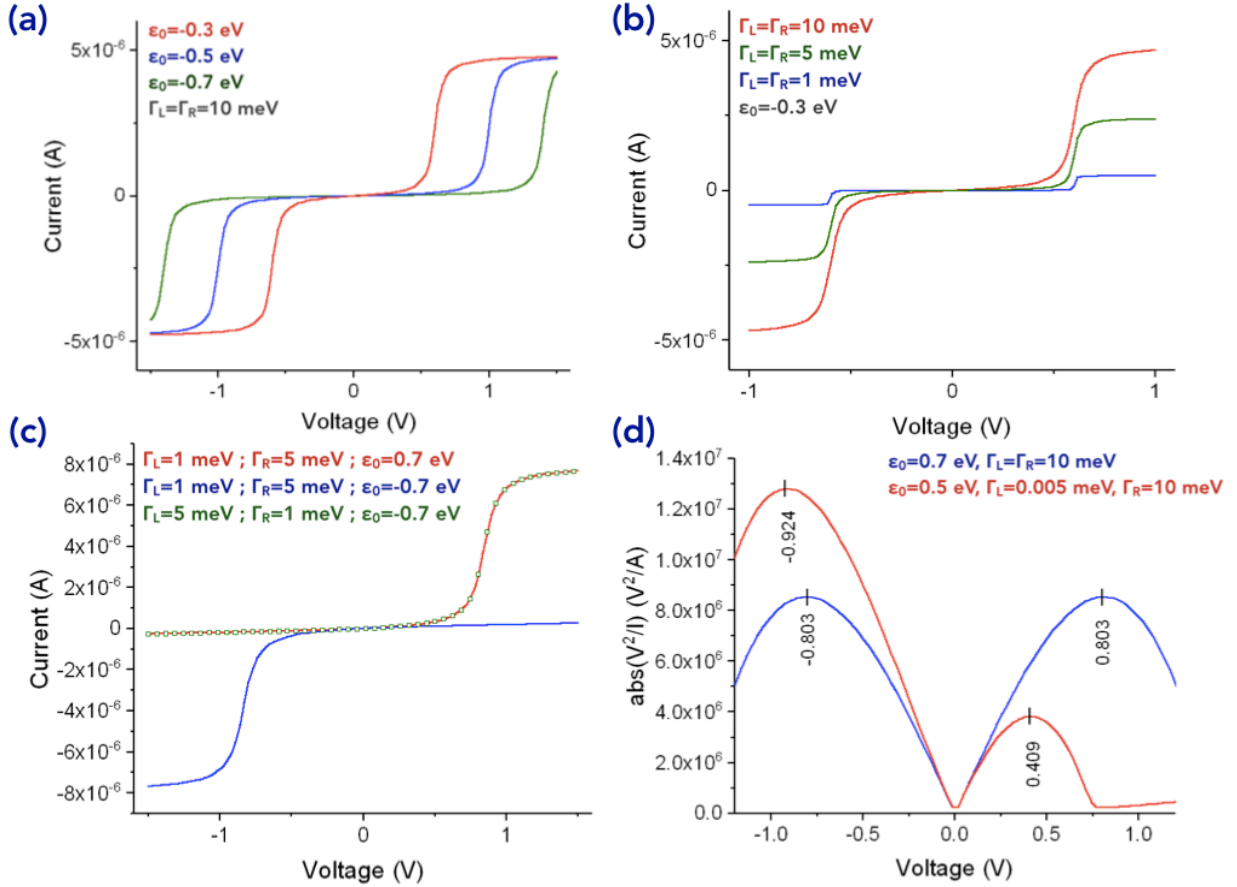

***Figure 4**. (a) Simulated I(V) curves for different values of the energy level $\varepsilon_0$ ($\Gamma_L=\Gamma_R=10$ meV). (b) Simulated I(V) curves for different values of coupling energy $\Gamma_L=\Gamma_R$ ($\varepsilon_0=0.3$ eV). (c) Simulated I(V) curves with $\Gamma_L\neq\Gamma_R$. (d) Transition voltage spectroscopy (TVS) : plot of abs($V^2$/I) vs. V and determination of the negative and positive threshold voltage ($V_{T-}$ and $V_{T+}$) for 2 sets of parameters. Red curve: $V_{T-}=-0924$V, $V_{T+}=0.409$V, given $\varepsilon_0=0.5$V with Eq. [5]. Blue curve: $V_{T-}=V_{T+}=0.803$V, given $\varepsilon_0=0.7$V with Eq. [5].*

However, this approach is subject to the same cautions as the one energy-level model since Eq [5] and Eq. [4] are based on the same approximations. In addition, energy levels associated to defects or impurities at the molecule/electrode interfaces (e.g., silicon and metal electrodes covered by ultra-thin native oxides) can be detected and wrongly attributed to the molecule.[60] In conclusion, these analysis of the experimental I(V)s curves have to be considered as qualitative and must be carefully checked versus ab-initio calculations.

Inside molecular junctions with more than a single molecule, the classical electrical circuitry laws are no longer valid, *i.e.,* the conductance of N molecules associated in parallel is no longer the sum of the conductance of each individual molecule. This feature is due to molecule-molecule interactions, *e.g.,* π-π molecular interactions.[61, 62] Reuter *et al.*[63] theoretically predicted that these molecule-molecule interactions should induce an asymmetry of the conductance histogram distribution (no longer Gaussian). This prediction was recently verified by varying the density of interacting molecules in nanodot/molecule junctions (NMJs) connected by conductive AFM.[64] In these experiments (Fig. 6), a classical log-normal distribution was observed for diluted molecules (weak interaction) – Fig. 6d. On the contrary, the conductance histograms for densely packed molecules (strong interactions) clearly exhibited an asymmetric distribution with a tail towards the low conductance values (red arrow, Fig. 6c) as a fingerprint of these interactions.[63] From this distribution, we determined the π-π interaction energy (30-35 meV) in good agreement with first-principle calculations.[64]

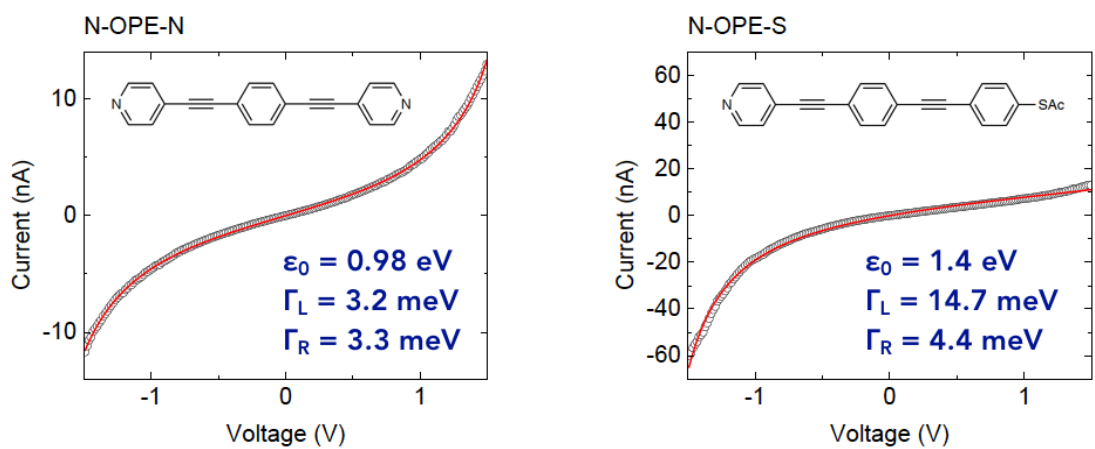

*Figure 5. Typical examples of the fits of the one energy-level model on molecular junctions with symmetric anchoring groups N-OPE-N and asymmetric anchoring groups N-OPE-S (OPE : olygo(phenylene ethynylene) ; N and S refer to pyridine and thiol anchoring groups, respectively). The single molecule I(V) curves are measured by MCBJ (mechanically controlled break junction).[52]*

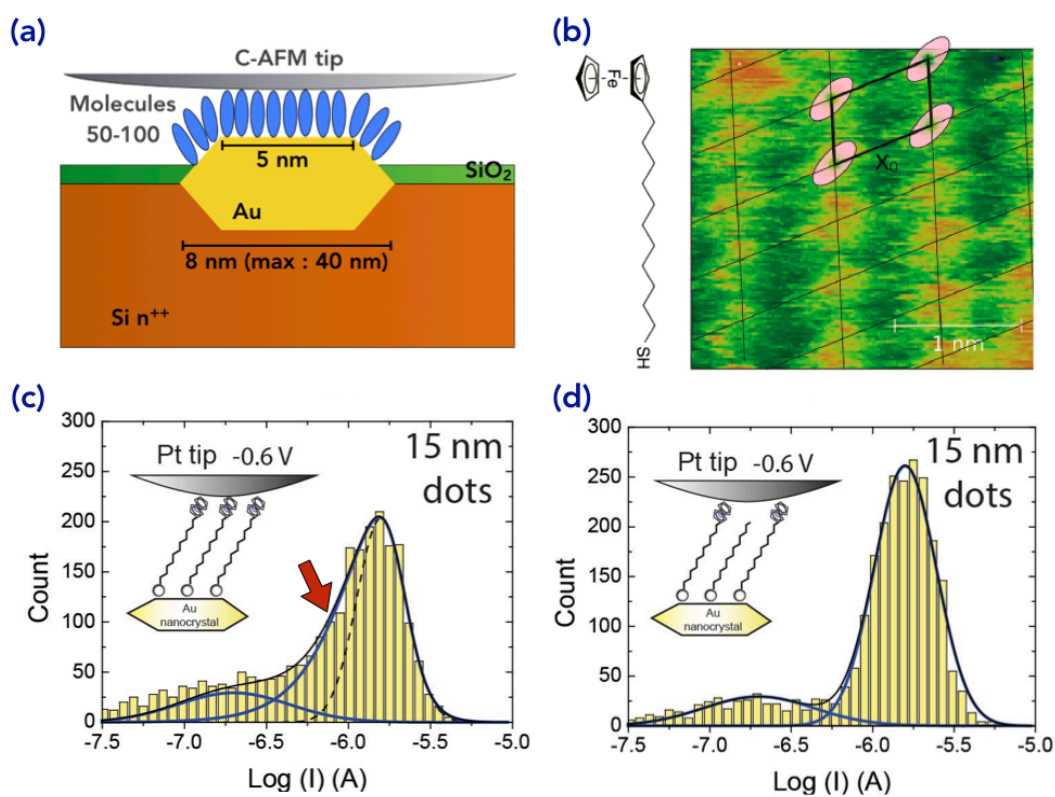

*Figure 6. (a) Scheme (cross-section) of the NMJ (nanodot-molecules-junction). Tiny gold nanodots (diameter 5-40 nm) are fabricated (e-beam lithography) on highly doped Si substrate and covered by a monolayer of chemisorbed ferrocenyl-undecanethiol molecules. (b) Chemical structure of the molecule (Fc-C11-SH) and high-resolution image of the ferrocenyl moieties (orange spot) organized on the top flat surface of the Au nanodot measured by STM in UHV. (c) Current histogram acquired by C-AFM on ~ 3000 NMJs for a densely packed monolayer (to favour the intermolecular π-π interactions), and (d) for a diluted monolayer (~ 10 alkylthiol chains per ferrocenyl-undecanethiol).[64]*

## 5. Electron transport at high frequencies

The typical electron transit time though a molecular junction considered as a tunnel barrier (off-resonant transport) is given by:

$$\tau = \sqrt{\frac{m}{2\Delta}} d \qquad [6]$$

with m the electron mass, d the molecule length and $\Delta$ the tunnel barrier (i.e. $\varepsilon_L-\varepsilon_F$ or $\varepsilon_F-\varepsilon_H$), predicting transit times of few femtoseconds considering d=1 nm and $\Delta$=1 eV and in the ps regime for resonant transport.[65, 66] Thus, operations of ME devices in the THz regime seems theoretically possible. However, up to now, molecular electronics devices were characterized in the DC regime, or at frequencies below MHz. We have recently demonstrated a molecular diode working at a frequency of 18 GHz with an estimation of the cut-off frequency of 520 GHz.[67] Such a value is on a par with the performance of RF-silicon Schottky diodes. These results were obtained by combining a tiny molecular diode made of about few tens of ferrocenyl-alkylthiol molecules chemically grafted on a nanodot Au electrode (5-20 nm in diameter) and connected with the tip of a home-made modified interferometer scanning microwave microscope (iSMM) measuring simultaneously the DC current and the microwave reflection signal $S_{11}$.[68, 69] Figure 7 shows the diode rectification behavior at DC, ~4 and 18 GHz. We clearly observe a rectification behavior of ~12 dB of the measured microwave reflection signal $S_{11}$ at 4 GHz (~4 dB at 18 GHz). The dynamic conductance vs. voltage curves at 18 GHz (deduced from the $S_{11}$ measurements) and the DC conductance (measured simultaneously) are similar (Fig. 6d), demonstrating that the molecular rectification behavior is preserved up to 18 GHz. These results demonstrate that molecular electronics is prone to high-frequency operation, and open perspectives to explore theoretically predicted exotic effects such as conductance amplification due to dynamic resonance of electron transport [70, 71] and THz molecular switches.[72]

## 6. Spin-dependent electron transport in molecular junctions

Using the spin magnetic moment of the electron instead of its charge to encode and process information (spin electronics, or spintronics) is a field of research increasingly studied. This spin can only have two quantum states (+ ½, we speak of "spin up" or - ½ "spin down"). This binary system thus allows encoding a logical "bit" "0 or 1". For organic materials and molecules, the weak spin-orbit coupling of carbon and the weak hyperfine interaction of the electron spins with the nucleus therefore suggests longer spin lifetimes (time during which the spin can remain polarized "up" or "down" before reversing due to various interactions with its environment). Tunnel magnetoresistance (TMR) through a ferromagnetic metal (FM)-molecular monolayer-FM junction was first reported in 2004.[73] The principle is identical to that of inorganic spin-valve junctions, the tunnel insulator is here played by a monolayer of alkyl chains, sandwiched between two ferromagnetic (Ni) electrodes. A TMR of 16% (at 4K) has been observed, but with a strong dispersion from one sample to another (e.g., positive and negative TMR have been observed for the same system). The presence of localized defects in the molecular tunnel barrier could be responsible for these dispersion. A correlation between the TMR decay and the diffraction of spins at the metal-molecule interface by molecular vibration modes was suggested on Ni-octanethiol-Co junctions.[74] These results were rationalized and the TMR performances improved by a detailed study of the spin-dependent hybridization at FM electrode/molecule interface, demonstrating that the spin polarization of the molecule-functionalized FM electrodes can be inverted or enhanced depending on the spin-dependent coupling strength between the molecules and the FM electrodes,[75] Fig. 8a. This has led to the birth of a new research field on molecular interfaces, dubbed "spinterface",[76-78] with many confirmations spanning from STM (Fig. 8b) to photoemission experiments.[79-85] Progress was obtained recently in the fabrication of molecular junctions on FM electrodes, for example by the chemical grafting of alkylphosphonic derivatives monolayers on $La_{0.7}Sr_{0.3}MnO_3$ [86-88] with reported TMR up to $10^4$ % (Fig. 8c) and with a significant stability up to high voltages (few volts on nm-thick monolayers).[89, 90] Simple molecules (alkyl chains) were used in these works and more functional (or stimuli responsive) molecules are mandatory for more elaborated molecular devices [39, 91] e.g., redox molecules for memory, photochromes for electro-optical molecular devices.

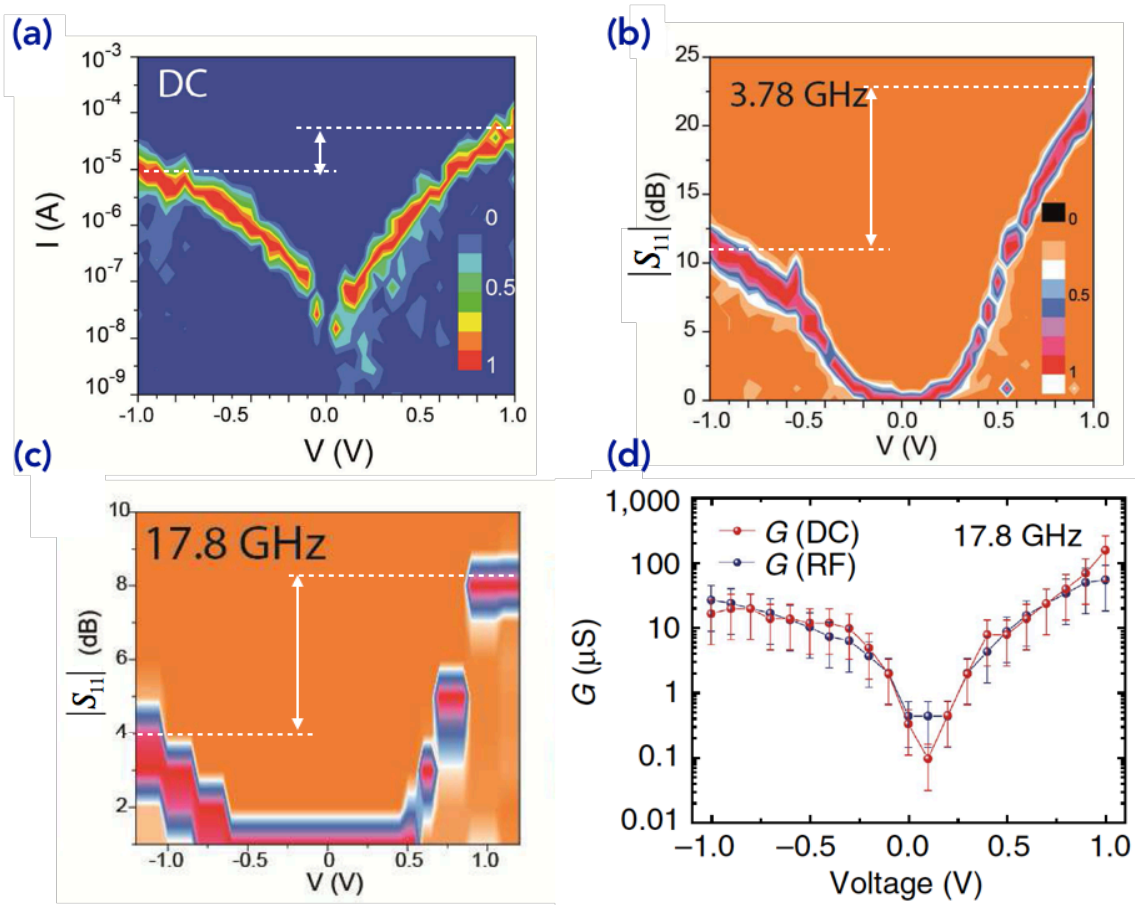

*Figure 7*. Histograms ("heat map") of (a) the DC current versus DC voltage acquired on a network of 100 ferrocenyl-undecanethiol molecular junctions (NMJs, Fig. 5a) and the simultaneously measured amplitude of the microwave $S_{11}$ parameter at (b) 3.8 GHz and (c) 18 GHz. (d) Comparison of the DC conductance and the microwave conductance at 18 GHz (extracted from the $S_{11}$-V data).[67]

Beyond this first step, we should expect that the possibility of tuning the resistance as well as the magnetoresistance thanks to the spin-dependent hybridization at the FM electrode/molecule interface could lead to a new class of devices combining analogical and digital properties. When photoswitch molecules are used with FM electrodes, the spin-polarized electron transport through the FM/molecules/FM junctions will depend on the conformation of the molecules and the molecule/electrode atomic contact geometry as evaluated from theoretical studies.[92-94] However, the chemical grafting of monolayers of functional molecules (more complex structures than simple alkyl chains) on FM electrodes remain challenging. We have recently reported the optically induced conductance switching at the nanoscale (conductive-AFM) of diarylethene derivatives self-assembled monolayers (SAMs) on $La_{0.7}Sr_{0.3}MnO_3$ electrodes,[95] and observed a weak conductance switching of the diarylethene molecular junctions (*closed* isomer/*open* isomer conductance ratios $R_{c/o}$<10), partly hidden under some conditions by the optically induced conductance switching of the $La_{0.7}Sr_{0.3}MnO_3$ substrate, and conductance ratio (*cis*/*trans* isomers) of about 20 for azobenzene derivatives on Co [48] - Fig 8d. These performances are lower than those of molecular junctions of the same azobenzene derivatives on gold electrodes for which conductance ratio up to ≈ 7x10³ were measured [96] and $R_{c/o}$ ≈ 100 were calculated and measured for diarylethene derivatives.[97, 98] These results call to more experimental and theoretical works to design stimuli-responsive molecular spin valves with higher performances.

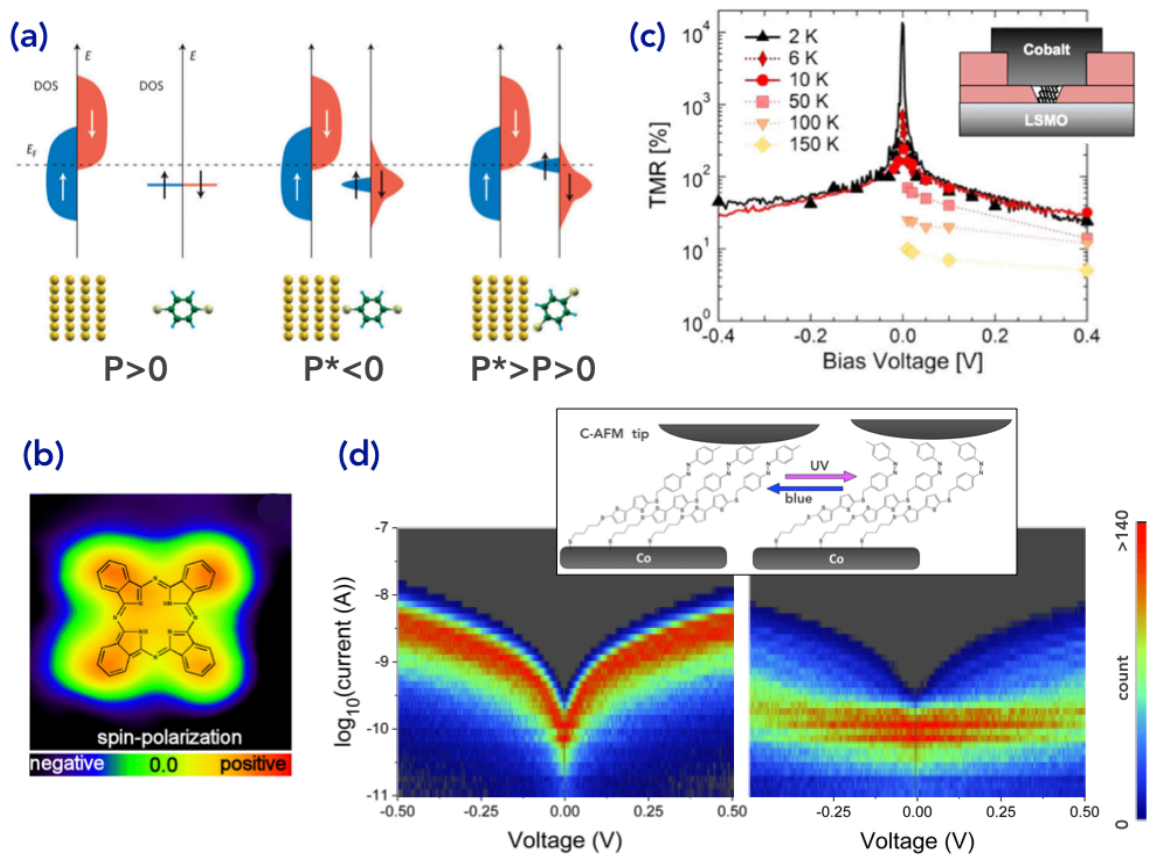

*Figure 8*. (a) Scheme of the spinterface concept. A ferromagnetic electrode with a positive spin polarization (higher density of spin-up states) is functionalized with a molecule. Depending on the shift and broadening of the molecular orbitals, the spin polarization P* of the functionalized electrode can be inverted (P*<0) or enhanced (P*>P>0). [99] (b) Experimental example of the spinterface effect. Spin-polarized STM image of a single metal-free phthalocyanine molecule (H2Pc) on a Fe surface showing the spin polarization inversion with respect to the Fe surface. [80] (c) TMR up $10^4$ (at 2K) measured in a LSMO/tetradecane phosphonic acid monolayer/Co spin valve. [90] (d) Current-voltage 2D histograms of Co/ azobenzene-bithiophene-alkylthiol monolayer/C-AFM tip for the "trans" and "cis" conformation of the molecules. The conformational switch is triggered by light (UV light for trans-to-cis, and visible light for cis-to-trans). [48]

Another approach to molecular spintronics consists in using one (or more) magnetic molecule in a molecular junction. The first experiments were performed with paramagnetic molecules (terpyridinyl lingands) complexing a magnetic ion (Co).[100] In a regime of strong coupling of the molecule with the electrodes and at low temperature (<25-30K), a Kondo effect, characteristic of the presence of an unpaired spin electron, has been observed. In a transistor configuration (with a bottom gate electrode), and with a molecule containing two magnetic centers (two vanadium atoms) it has been shown that this effect can be inhibited or activated by playing on the charge state of the molecule by applying a gate voltage (switching between the Coulomb blockade regime and the Kondo regime).[101] These first demonstrations of a molecular spin transistor opened the door to other possibilities such as the magnetic molecular spin valve (a magnetic molecule between two ferromagnetic electrodes) – see a review in [102] or the implementation of spin qubits in a magnetic molecule electrically controlled by an STM tip for example.[103] It is also possible to use the spin state of a molecule attached to a non-magnetic one-dimensional conductor (e.g., a carbon nanotube) to modulate the electronic transport in this conductor. This modulation could be detected directly by measuring the current, or by using an ultra-sensitive carbon nanotube nano-squid that allows measuring small variations of magnetic flux.[104] The topics of Kondo physics and molecular magnet electronics have been observed and studied in a wide variety of molecules, the reader is referred to recent review papers for more details.[105-107]

## 7. Molecular electronic plasmonics

In nanophotonics, a promising approach is devices based on plasmons – oscillations of electrons at metal-dielectric interfaces – which can operate at optical frequencies. Unlike photons, plasmons can be manipulated with metal nanostructures of a few tens of nanometers, i.e., far below the diffraction limit. The ability to capture and confine light in nanoscale structures below 100 nm in the form of so-called surface plasmons (SPs) may transform current technologies. SPs confine and enhance local electromagnetic fields near surfaces of metallic nanostructures at optical frequencies and have the ability to propagate along sub-diffractive metallic waveguide opening-up new perspectives for integrated opto-electronic circuits at the nanoscale.[108-112] On the other hand, molecular-scale electronics operates at the length scale of few nanometers, thus combining molecular devices and plasmonics is an opportunity to study electron/plasmon interaction in the quantum regime.

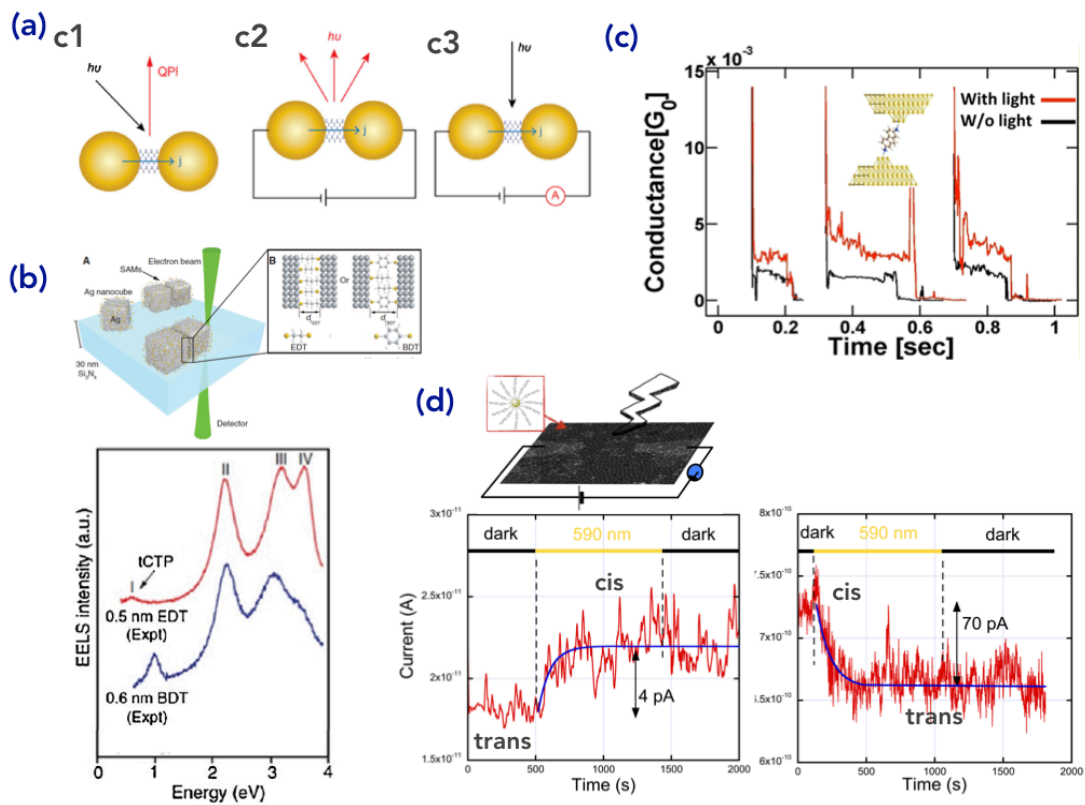

*Figure 9*. (a) Schemes of operation principles for molecular electronic plasmonics.[113] (b) Direct observation of the tunnel charge transfer plasmon (tCTP) in Ag nanocube dimers connected by 2 types of molecules.[114] (c) Plasmon-assisted tunneling through a single diaminofluorene molecule junction.[115] (d) Plasmon-induced isomerization of azobenzene derivatives and corresponding variation of electron transport properties of a 2D network of Au nanoparticles (10 nm diameter) capped with azobenzene bithiophene molecules.[116]

Figure 9a shows several possible ways how molecular tunneling junctions based on electrode-molecules-electrode configurations can be combined with plasmonics. For simplicity the electrodes are drawn as spherical plasmonic resonators, but other forms (rods, planar surface…) are possible. The molecular component can be present in the form of a single molecule, or a self-assembled monolayer (SAM). In the configuration c1 (Fig. 9a), molecular junctions are used to study quantum plasmonics. Two closely spaced plasmonic resonators are bridged by a SAM onto which plasmons are excited. Usually this is done by incident light or by an electron beam inside a transmission electron microscope. These plasmons induce an electric field across the gap resulting in quantum mechanical tunneling across the molecules leading to quantum plasmon resonances (QPR) such as the so-called charge transfer plasmon (CTP) modes.[117, 118] In this dimer configuration, if the gap is below < ~1 nm, a new quantum plasmon, the tunneling charge transfer plasmon mode (tCTP) has been observed in junctions consisting of two silver nanocubes separated by SAM of 1,2-ethanedithiolates (EDT) or 1,4-benzenedithiolates (BDT) (Figure 9b). The emission energy of the tCTP can be tuned by changing the molecular structures (and by doing so the tunneling barrier height) bridging two plasmonic resonators.

In the configuration c2 (Fig. 9a), a voltage is applied across the molecular gap. The tunneling charge carriers then excite plasmon modes in the electrode material, either directly by the tunneling charge carriers or via electroluminescence from the molecules inside the junction.[119] In this molecular electronic plasmon excitation devices, the properties of the generated plasmons can be controlled molecular electronically (without the need for optical nanoantennas): i) the polarization of the plasmon depends on the tilt angle of the SAM, ii) the frequency of the plasmon depends on the applied bias, and iii) the bias-selective plasmon excitation in only one direction of the bias using a molecular diode.[112]

The reverse process, i.e., the coupling of a plasmon to tunneling charge carriers, can also happen.[120] In this configuration (c3, Fig. 9a), the plasmons are excited in the junction via an external light source and they couple to the tunneling charge carriers and increase the tunneling current across the junction (so-called optical rectification). In this case, molecular electronics is applied to detect plasmons. A first mechanism is plasmon-assisted tunneling (PAT) because the plasmon field modulates the tunnel barrier in the junction. A second, indirect, mechanism is possible if the molecules absorb the plasmon energy and generate electron-hole pairs. A typical example of PAT is shown in Fig. 9c for a single molecule (diaminofluorene) junction.[115] The plasmon field enhancement in the molecular junction is about $10^3$. Finally, if it is known that the redox and/or conformational states of molecules linked to plasmonic nanostructures can shift the SP frequency.[121-123] It has been recently demonstrated that SP can also be used to induce the isomerization of molecular switches (i.e., azobenzene derivatives) and then a significant change of the current in the devices (Fig. 9d).[116] A plasmon-induced resonance energy transfer (PIRET) mechanism [124] is likely responsible for this effect. This plasmon-induced isomerization (PII) is faster (about a factor 10) than the usual isomerization triggered by UV-visible light, which may be helpful for light-driven molecular memory and light reconfigurable molecular circuits.

A detailed and comprehensive review on molecular electronic plasmonic has been recently published in [113].

## 8. Quantum interference and thermal transport

Thermoelectricity at the nanoscale is based on the seminal work of Hicks and Dresselhaus who have established theoretically that one-dimensional quantum wires (formed by adjacent metal atoms) should lead to highly efficient thermoelectric systems compared to 2D and 3D systems.[125] They demonstrated that lowering the dimensionality of the systems improves the electronic quantities that govern thermoelectricity. In molecular junction, the first theoretical analysis of the thermoelectric properties was reported by [126] for a benzenedithiol molecule. Since that, many works have shown theoretically and experimentally that molecular junctions indeed exhibit very interesting thermoelectric characteristics; they are described in a series of review papers.[106, 127-129]

The thermoelectric properties are described by 3 simple equations:

$$S = -\frac{\Delta V}{\Delta T} \quad [7] \; ; \quad ZT = \frac{\sigma S^2}{\kappa} = \frac{G_{el} S^2}{G_{th}} \quad [8]; \quad PF = \sigma S^2 \; or \; G_{el} S^2 \quad [9]$$

where S is the Seebeck coefficient, or thermopower ($\Delta V$ being the difference of voltage generated by applying a difference of temperature $\Delta T$ between the electrodes – Fig. 10). The efficiency of energy conversion is evaluated by the thermoelectric figure of merit ZT. This metric is used whatever the materials and systems used to fabricate the thermoelectric devices and can be used for comparison: the higher is ZT, the best is the material. In these equations, $\sigma$ and $\kappa$ are the electrical and thermal conductivity, respectively, or equivalently at the molecular level (single and few molecules devices) where dimensions are ill-defined, the electronic and thermal conductance $G_{el}$ and $G_{th}$, respectively. PF is the power factor. The thermal conductivity/conductance is the sum of a vibrational contribution and an electronic contribution, $\kappa = \kappa_v + \kappa_e$ ($G_{th} = G_{th,v} + G_{th,e}$). The vibrational contribution is negligible at low temperature, but it increases with temperature and can eventually dominate the electronic contribution (this latter being significant if the molecule/material is enough conducting). However, in molecular junction, the electronic contribution is usually negligible with $G_{th,v}$ of the order of 1-70 pW.K$^{-1}$.[128, 129]

The Seebeck coefficient can be described within the Landauer formalism:

$$S = -\frac{\pi^2 k_B^2 T}{3e} \frac{\partial \ln T(E)}{\partial E}\bigg|_{E=\varepsilon_F} = -\frac{\pi^2 k_B^2 T}{3e} \frac{1}{T(\varepsilon_F)} \frac{\partial T(E)}{\partial E}\bigg|_{E=\varepsilon_F} \quad [10]$$

with T(E) the transmission probability of electrons through the molecular junction. Since the sign of S is related to the slope of T(E), it is a fingerprint whether the charge transport though the molecular junction occurs via the LUMO or the HOMO, i.e., it depends on the direction of the electron flow (Fig. 10a), it is positive if going to the hot side (thus promoted by HOMO, Fig. 10a), and negative the other way around, i.e., electron transport through the LUMO (Fig. 10b). Experimental and theoretical values of S ranges from 0.5 to 30 µV.K$^{-1}$.[128, 129]

From the analysis of the known values of S, $G_{el}$ and $G_{th}$ of a large variety of molecular junction, it may be inferred that molecular junctions are prone for a high thermoelectric capacity with expected ZT higher than in bulk systems (values of 3 to 4 have been theoretically predicted).[130] However this requires that high values of S and $G_{el}$ and a low value of $G_{th}$ are obtained for the same molecule, a remaining challenge in molecular electronics. For instance, the electronic contribution to the thermal conductance, $G_{th,el}$ is a function of the electronic conductance $G_{el}$, thus these parameters cannot be tuned independently.

Several factors can strongly influence T(E) and its derivative, and thus the thermoelectric properties, Fig. 10. First, the position of the HOMO/LUMO levels with respect of the Fermi energy. Moving these levels close to $\varepsilon_F$ increase the electronic conductance $G_{el}$ and the slope $\partial \ln(T(E))/\partial E$ (thus the Seebeck coefficient S) – Fig. 10c. Second, a decrease of the coupling energies ($\Gamma_L$ and $\Gamma_R$) increases the slope of ln(T(E)) at $\varepsilon_F$ but it decreases the conductance (see Eq. 4), Fig. 10d, thus at the risk of decreasing the power factor PF (Eq. 9). Third, quantum interferences (QI) have been theoretically proposed [131] to improve S and ZT by introducing narrower and asymmetry peaks in T(E) around the Fermi level, Fig. 10e.

Quantum interferences in molecular junctions [61, 132] can arise in several situations. The first one occurs if the electron transmission is mediated by a combination of different molecular orbitals (Fig. 11a). The quantum mechanics wavefunctions (described by amplitude and phase) of the electrons passing through one or another orbital may have their phases similar or in opposition (phase difference of π) resulting to constructive or destructive interferences, respectively, that increase of decrease the conductance (a dip is created in T(E) near $\varepsilon_F$). The typical example is the benzene-based molecular junctions with connections to the electrodes in para or meta positions (Fig. 11b), and several experimental results were reported for several π-conjugated short molecules.[133-139]

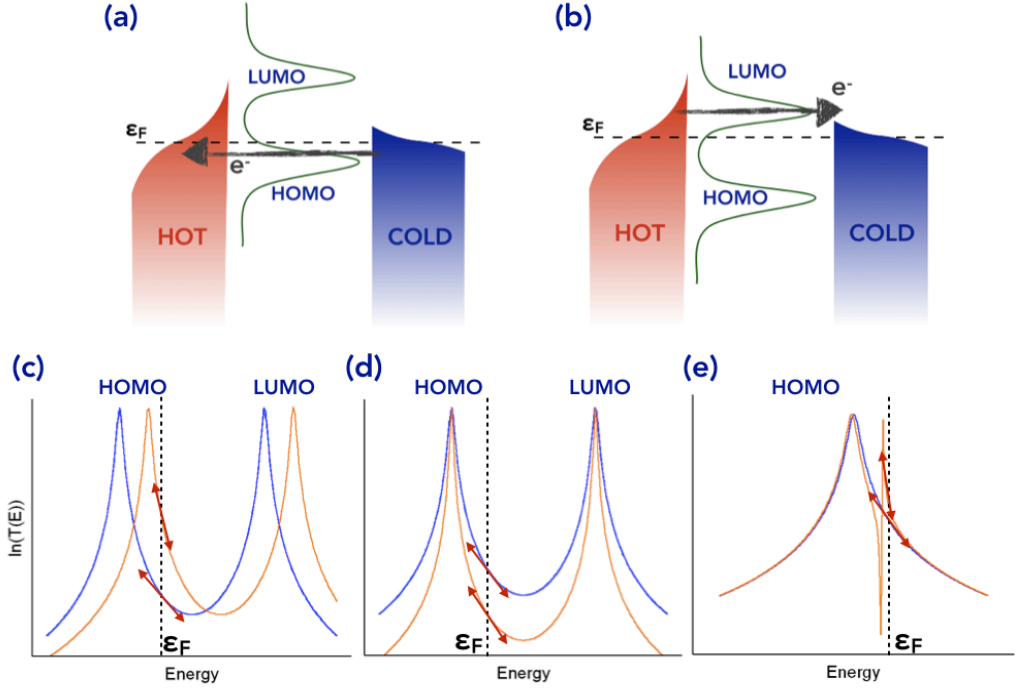

*Figure 10. Energy scheme of a molecular junction submitted to a temperature gradient (shown by the red and blue broadened Fermi-Dirac functions in the electrodes). When the HOMO is close to the Fermi energy (a), the net flow of electrons goes to the hot electrodes, while it goes to the cold electrode if the LUMO is close to the Fermi energy (b). Typical transmission coefficient T(E) and slope ∂ln(T(E))/∂E (red arrows) illustrating three ways to improve the Seebeck coefficient: (c) shift of the molecular orbitals closer to $\varepsilon_F$, (d) reducing the electrode coupling energy Γ and (c) introducing a sharp resonance (here a Fano resonance) near the Fermi energy. In (c) and (d) the blue lines are simulations of Eq. (1.3) with $\varepsilon_0$ = -0.5/0.5 eV (HOMO/LUMO), $\Gamma_L=\Gamma_R$ = 10 meV; the orange line in (c) with $\varepsilon_0$ = -0.3/0.7 eV (HOMO/LUMO), $\Gamma_L=\Gamma_R$ = 10 meV; the orange in (d) with $\varepsilon_0$ = -0.5/0.5 eV (HOMO/LUMO), $\Gamma_L=\Gamma_R$ = 5 meV. In (e) Fano resonance Eq. 11 with $\varepsilon_0$ = -0.5eV, Γ = 10 meV, $\varepsilon_1$ = -0.4 and u = 5 meV (orange curve) and without (blue curve) Fano resonance ($\varepsilon_1$ and u = 0).*

Another possible QI is the destructive Fano resonance,[140] which occurs if, near the Fermi energy, the molecular junction has a delocalized molecular orbital between the two electrodes and a discrete energy level localized on another part of the molecule (Fig. 11c), for example due to a chemical moiety on a side of the molecule backbone (i.e., a pendant group). If these two energy levels are close in energy and the localized sate is coupled with the delocalized one but not directly to the electrodes, charge transport resonances will occur because of quantum confinement of electrons and T(E) will display an asymmetric Fano resonance/anti-resonance, resulting in a pronounced variation of its derivative and thus an enhancement of the thermopower (Fig. 10e). In this case, T(E) is now longer described by Eq. (3) but it writes:

$$T(E) = \frac{4\Gamma_L\Gamma_R}{\left(E-\varepsilon_0-\dfrac{u^2}{E-\varepsilon_1}\right)^2 + \left(\Gamma_L+\Gamma_R\right)^2} \quad [11]$$

with $\varepsilon_0$ and $\varepsilon_1$ the energy level of the delocalized and localized states, respectively, and $\Gamma_L$ and $\Gamma_R$ the coupling energy of the delocalized states to the electrodes and u the coupling energy between the two states (Fig. 11c).

Experimental and theoretical results have been published evoking Fano resonance to explain the electron transport of specifically designed molecular junctions.[141-151] Among them, a typical example concerns anthraquinone derivatives that are cross-conjugated molecules showing a lower conductance compared to their linearly conjugated counterparts (Fig. 11d). The cross-conjugated anthraquinone displays a pronounced dip around 0 bias that is the fingerprint of a destructive Fano anti-resonance.[142, 144, 147, 152]

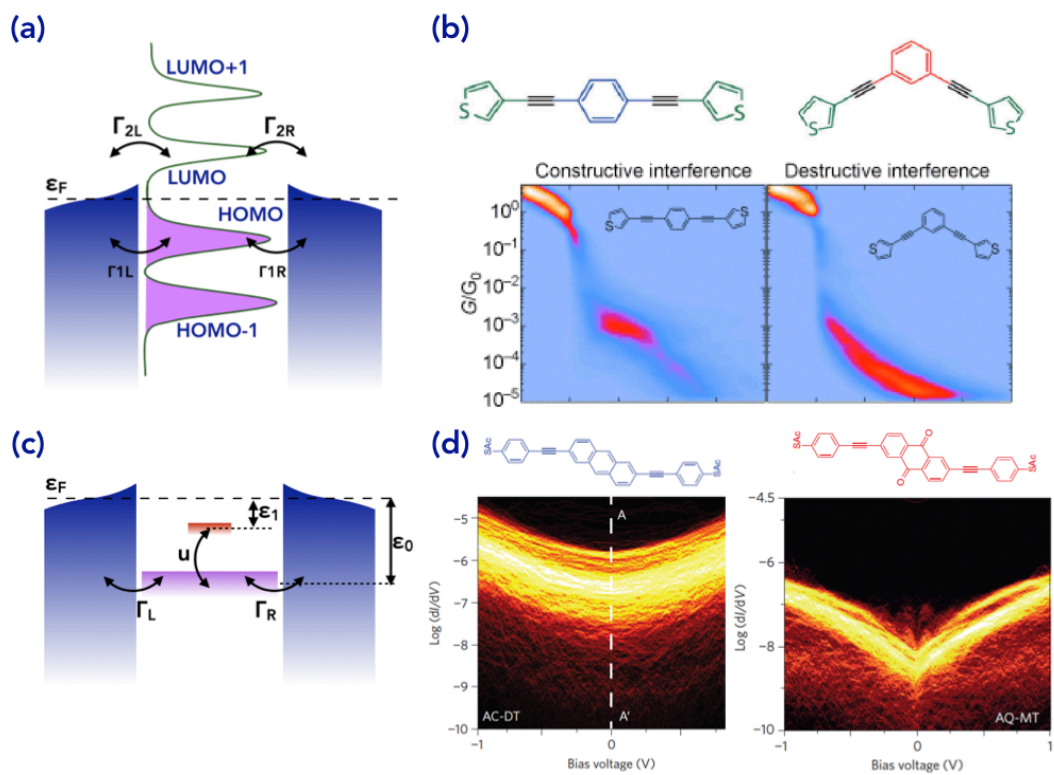

*Figure 11. (a) Quantum interference energy scheme when two energy levels (here HOMO and LUMO) are combined in the electron transmission. (b) Typical examples of constructive and destructive QIs in a benzene derivative molecule connected to the electrodes in the para and meta position, respectively. The MCBJ measurements show a plateau at a higher conductance for the molecule in the para position.[135] (c) Energy scheme for a molecular junction with a Fano resonance between a delocalized state (purple) and a localized state (red) coupled with the delocalized state but not to the electrodes. (d) Comparison of the conductance-voltage curves (conductive-AFM measurements) of a cross-conjugated anthraquinone and a linearly conjugated analog. The anthraquinone molecular junction shows a lower current with a marked dip in the conductance at 0 bias.[142]*

Albeit ZT values as high as theoretically predicted (ZT ≈ 4 at room temperature [130]) are not yet experimentally demonstrated using this QI approach, these findings pave the way towards important improvements and higher level of sophistication of the thermoelectric molecular devices by adapting its intrinsic components (nature of electrodes, molecular backbone, and interfacial coupling) to obtain the required efficiency at working conditions.

## 9. Noise in molecular junctions

Noise (stochastic fluctuations of a signal or quantity) is a ubiquitous phenomenon in nature and the low-frequency noise (LFN) or the so-called 1/f noise (also known as flicker noise, discovered by Johnson [153] in vacuum tubes) is observed in many fields (physics, astrophysics, biology, economics, technology…), not only in electronic devices.[154] In molecular electronics, noise spectroscopy has been studied this last decade (or so) to reveal subtle electron transport phenomena not observed using conventional DC current measurements.

Noise in electronic devices is characterized by its frequency dependence according to its physical origin. Basically, the time-dependent current in any electronic device is written as $I(t)=\langle I \rangle + \delta I(t)$ with $\langle I \rangle$ the average current and $\delta I(t)$ the time-dependent fluctuations. Due to the stochastic nature of noise, the noise is more conveniently characterized in the frequency domain, via Fourier transformation, using its power spectral density (PSD), $S_I(f)=\langle\delta I(t)^2\rangle/\Delta f$ where $\langle\delta I(t)^2\rangle$ is the variance measured at a frequency f over a bandwidth $\Delta f$.[155] The most basic noise is the thermal noise (or Johnson-Nyquist noise), which is due to the random thermal motion of electrons in the devices or materials and exits in any of them. This noise is independent of the frequency (white noise) and given by a current PSD $S_I=4k_BT/R$ with $k_B$ the Boltzmann constant, T the temperature and R the resistance. It is usually the noise floor of the device. In the following, I briefly describe some recent results on 1/f noise, RTS (random telegraph signal) noise and shot noise in molecular junctions. For more details, see a recently published review.[156]

1/f noise refers to any fluctuations with a PSD proportional to $1/f^n$ with n usually close to 1, but more generally $0.5 \lesssim n \lesssim 2$, the value of n depending on the physical origin of the noise. In bulk materials and devices, 1/f noise is related to the number fluctuation or mobility fluctuation phenomena (see a review in [157]). The first one corresponds to the fluctuations of the number of charge carriers in the devices, which can arise due to trapping/detrapping of charges by traps (i.e. defects) in the materials and/or at interfaces.[158] In the second one,[159] the mobility of charges fluctuates due to scattering by various events (impurities, defects, electron-phonon interactions,…). Note that both mechanisms can be active simultaneously. The current noise PSD is then strictly proportional to 1/f and scales as $I^2$ with I the DC current.[157] Such a $1/f^n$ noise was observed (Fig. 12a) in molecular junctions, silicon-alkyl chain $(C_8H_{37})$-Al, with 1<n<1.2 (n slightly increasing with the applied voltage).[160] The normalized current PSD $(S_I/I^2)$ reveals a bump at voltages between 0.4 – 1V (Fig. 12b), which was attributed to localized states (localized energy levels associated to defects/impurities) into the alkyl tunnel barrier. This behavior was consistently modeled by an energy-dependent trap-induced tunnel current and considering the effect of traping/detrapping of electrons by a two-level system (Fig. 12b).[161, 162] This model also predicts that the current PSD, $S_I$, scales as $(\partial I/\partial V)^2$ as experimentally observed.[160] The bump in the $S_I/I^2$ versus V is sample dependent (Fig. 12b) and it is likely due to an energy-peaked (modeled by a Gaussian) distribution of energy states in resonance with the Fermi energy windows for V>0.4 V. The same result (bump in the normalized PSD) was latter observed in Au-hexanedithiol-Au junctions and similarly ascribed to localizes states (defects) in the molecular junctions.[163] It is likely that these localized states are induced by the top metal (large area) deposition on the molecular monolayer in these cases. At the nanoscale, noise measurements on SAMs of alkylthiols and short π-conjugated molecules (4-mercaptopyridine, benzenedithiol) on Au by C-AFM showed that the current PSD is proportional to $1/f^2$.[164, 165] This behavior was attributed to molecule-electrode interface bond (i.e., S-Au bonds) fluctuations that induce conductance fluctuations in the molecular junctions.

In tiny devices or for a small number of noise sources, RTS noise is observed in the time domain. RTS noise manifests as abrupt, discrete, current switching between two distinct levels, Figs. 12c and d (albeit more levels are also observed) and it has been widely observed and characterized in nanoelectronic devices.[166] Figure 12c shows the first observation of RTS noise in a single (or a few) molecular junctions.[167] In this experiment, a single or a few oligo-phenylene-ethynylene (OPE) derivatives were inserted in a monolayer of insulating molecules (alkylthiols) chemisorbed on Au surface and current images at a fixed bias were acquired by STM over time. The STM images showed a stochastic two-level switching of conductance of the OPE derivatives (Fig. 12c). This behavior was explained by the fluctuations of the S-Au bonds at the molecule/metal interface, the S-Au bond being randomly broken/reformed or the degree of hybridization between the thiol group and the metal can change, thus modulating the molecular junction conductance.[168] More recently, RTS noise was also observed in single molecule experiments (STM-BJ and MCBJ) for various molecules : di(methylthio)-stilbene, dithiadecane at low temperature (4_- 70K) by STM-BJ [169] and oligophenyle at room temperature by MCBJ [170] and MCBJ with alkanedithiol and oligo phenylene ethynylene [171]. Flicker noise was also observed at room temperature in the STM-BJ experiments with exponent n=1.7 and the PSD was found to scale with $(\partial I/\partial V)^2$ if the molecules are weakly coupled to the two electrodes and to about $(\partial I/\partial V)$ for a molecule more strongly coupled at the two sides.[169] For the oligophenyl MCBJ experiments, the PSD exhibits $1/f^2$ dependence as expected for RTS noise (Lorentzian PSD, $S_I \propto \tau/(1+(2\pi f\tau)^2)$ with τ the effective time constant of the RTS noise). In both experiments, the authors concluded that the RTS noise is due to the dynamic rearrangement of the molecule-metal bonds.

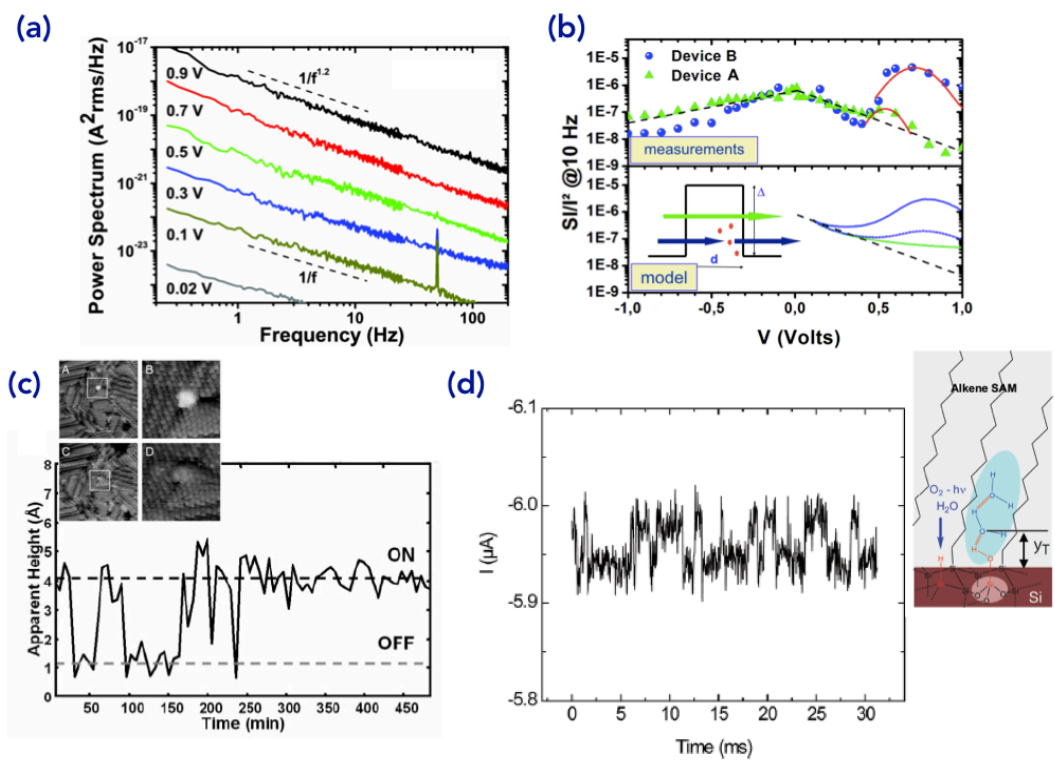

*Figure 12. (a) Current PSD, SI, in Si-alkyl-Al junctions at several voltage. (b) Normalized $SI/I^2$ for two Si-alkyl-Al junctions versus the applied voltage and scheme of the trap-assisted tunneling model explaining the bump at V>0.4V in the experiments, the bump amplitude being related to the density of defects in the molecular junctions.[160] (c) STM images showing a typical RTS noise (two-level fluctuations) observed for a single OPE molecules in a matrix of alkylthiol monolayer on Au.[167] (d) RTS noise due to an electrochemical reaction (redox) of water molecules trapped at the interface between silicon and alkyl molecules.[172]*

RTS noise was also used to study electrochemical reactions in the molecular junctions, such as redox reactions that can modify the transport mechanisms in the devices. In molecular junctions containing a low density of redox molecules (ferrocenyl-hexanethiol) diluted in a monolayer of non-redox molecules (alkylthiols), RTS noise was observed, while no RTS was detected in a pure non-redox molecular junction.[173] It was also demonstrated that a low density of water molecules trapped at the silicon-alkyle interface can induce RTS noise in the molecular junctions (Fig. 12d). The underlying mechanism was suggested to come from electrochemical charge transfer reaction between the silicon substrate and $H^+/H_2$ redox couple with a characteristic energy level close to the Fermi energy of the silicon electrode.[172]

Finally, shot noise (discovered by Schottky[174] in vacuum electronic tubes) is also an interesting phenomenon in molecular junctions. Shot noise is due to the discrete and quantized nature of electrons transmitted through the device. It is related to the transmission coefficient $T_m(E)$ where m denotes the number of transmission channels in the junction. At low temperature ($k_BT \ll eV$) and low transmission limit ($T_m \ll 1$), the shot noise PSD is simply $S_I = 2eVG_0\sum_m T_m = 2eI$ with $G_0$ the conductance quantum, V the applied voltage, and I the DC current.[175, 176] Shot noise was used to study the electron transport properties in single molecule junctions with deuterium ($D_2$) between Pt contacts,[177, 178] benzene,[179] benzenedithiol.[180, 181] These experiments demonstrated that the shot noise in the Pt-$D_2$-Pt junction is strongly reduced and that a single channel, with a high transmission probability, dominates the electron transport. Thus, in such a nearly fully transparent channel, the electrons are transmitted coherently in the molecular junctions and the shot noise vanishes.[177] In addition, it was found that, at higher voltages, shot noise spectroscopy could be used to detect molecular vibrations.[178] In a Pt-benzene-Pt molecular junction, where the benzene is connected to the electrodes by direct Pt-C bonds, the DC conductance is high (0.1 - $1G_0$). The shot noise measurements revealed that the electron transport occurs via several channels, the precise number depending on the molecule/electrode conformation.[179] For Au-benzenedithiol-Au junctions with a large range of conductance (0.01 - 0.24$G_0$), the shot noise measurements demonstrated that electrons are always transmitted through a single conduction channel. The authors concluded that the Au-S bonds at the interface is the limiting factor and that a direct tunneling from the Au contact to the benzene ring is hindered in that case.[180]

## 10. Conclusion and further reading

In this review, selected results are presented and discussed. More results and discussions are available in recent review papers. In addition to the review papers already cited in this chapter, several other review papers are of interest on ME in general [44, 91, 182, 183] and on more specific topics:
- ME on silicon and other semiconductor platforms,[16]
- molecular spintronics,[107, 184, 185]
- thermal transport in MJs,[129, 186]
- and focalized on theory and computational methods.[43, 65, 187]

In the perspective of applications, and although no ME devices are commercially available so far (and probably never will be) basic research in ME has significantly improved our fundamental understanding of physics at the nanoscale. Nevertheless, ME devices are envisioned to complement semiconductor devices by providing new functions or already existing functions at a simpler process level and at a lower cost by virtue of their self-organization capabilities. In addition to the general review papers mentioned above, significant results to implement electronic devices at the molecular-scale are reviewed on:
- molecular diodes,[11]
- molecular switches towards memory devices,[188-191]
- and more exploratory research on ME for unconventional computing such as neuromorphic computing and quantum computing can be found in [36, 103, 192-199].

## 11. Acknowledgements

I thank my colleagues at IEMN and outside the lab who contributed to works reported in this review, as well as financial supports from EU and ANR.